# Modulating Thermometric Performance via Dopant Concentration and Morphology in Luminescence Thermometer Exhibiting Dual Structural Phase Transitions

Malgorzata Kubicka[1], Maja Szymczak[1], Maciej Ptak[1], Damian Szymanski[1], Vasyl Kinzhybahlo[1], Marek Drozd[1], Lukasz Marciniak[1*]

[1] Institute of Low Temperature and Structure Research, Polish Academy of Sciences, Okólna 2, 50-422 Wrocław, Poland

*corresponding author: l.marciniak@intibs.pl

**Abstract**

Expanding the operational range of luminescent thermometers that utilize thermally induced structural phase transitions in lanthanide-doped materials necessitates the exploration of novel host matrices with diverse thermal behaviors. In line with this objective, the present study offers a comprehensive analysis of the temperature-dependent spectroscopic properties of $Li_3Sc_2(PO_4)_3$:$Eu^{3+}$. The findings reveal that the studied material undergoes two reversible phase transitions: $\gamma_{LT} \rightarrow \alpha/\beta$ phase transition at approximately 160 K, followed by an $\beta \rightarrow \gamma_{HT}$ transition around 550 K. These transitions are evidenced by notable alterations in the emission spectra and luminescence decay kinetics of $Eu^{3+}$ ions. By employing an appropriate luminescence intensity ratio, the sensitivity was determined to be 7.8 % $K^{-1}$ at 160 K for 0.1%$Eu^{3+}$ and 0.65 % $K^{-1}$ at 550 K for 0.5%$Eu^{3+}$. Furthermore, the study demonstrates that the phase transition temperature in $Li_3Sc_2(PO_4)_3$:$Eu^{3+}$ can be modulated through variations in dopant ion concentration and annealing conditions, which in turn influence the material's morphology. These strategies enable the fine-tuning of thermometric performance in phase transition-based luminescent thermometers. To the best of our knowledge, this represents the first report in the literature of a luminescent thermometer exhibiting dual thermal operating ranges.

**Introduction**

The high sensitivity of the spectroscopic properties of inorganic materials to temperature variations, as evidenced by changes in their emission spectra and luminescence kinetics, can be effectively utilized for remote temperature sensing[1–4]. Skillful exploitation of these variations enables rapid, precise, and reliable temperature measurements. The uncertainty in such measurements is inversely proportional to the sensor's thermal sensitivity, prompting intensive research aimed at maximizing relative sensitivity[2,5–7]. This research encompasses the development of new luminescent materials with optimal structural parameters-such as interionic distances and phonon energies-and the exploration of thermally induced mechanisms that alter phosphor spectroscopic properties[7–9]. These mechanisms include thermalization of energy levels, multi-phonon depopulation processes of excited states, and energy transfers between excited levels[10,11]. Among the various mechanisms employed in lanthanide ion-based luminescence thermometry, thermal coupling between two emitting states is particularly prominent[6,11]. When the energy gap between two or more levels is sufficiently small, thermal energy facilitates electron redistribution according to the Boltzmann distribution[12]. A key advantage of this approach is the theoretically predictable thermal variation in the population of emitting levels, and consequently, the luminescence intensity ratio of the bands associated with their radiative depopulation. However, the sensitivity of such temperature sensors is proportional to the energy separation between levels, which should not exceed 2000 cm$^{-1}$ to remain efficient thermal coupling[8]. As a result, these luminescence thermometers typically exhibit relative sensitivities not exceeding 2% K$^{-1}$ [6].

To achieve higher sensitivity values, alternative strategies are being explored. One promising approach involves modifying the spectroscopic properties of Ln$^{3+}$ ions through first-order structural phase transitions[13–20]. Changes in the point symmetry of the crystallographic sites occupied by lanthanide ions can alter the number of Stark components into which multiplets split, adjust the energy separations between them, and modify the probabilities of radiative depopulation of excited levels[17,19,20]. This enables the development of luminescent thermometers with exceptionally favorable performance characteristics, such as high relative sensitivity and extended luminescence lifetimes. However, due to the nature of the mechanisms inducing these spectroscopic changes, the operational temperature range of such thermometers is often quite narrow[16,19]. To broaden the operating temperature range, a recent strategy involves developing a series of luminescent thermometers by shifting the phase transition temperature[16]. One way to achieve this is the partial substitution of host material cations with optically inactive cations that have ionic radii different from those of the replaced ones. While this method has proven effective, it is often accompanied by a reduction in relative sensitivity[16,18]. Therefore, identifying new host materials that exhibit first-order phase transitions across various temperature ranges is a more advantageous approach.

In this study, we introduce a novel thermometric material that not only undergoes a phase transition at higher temperatures than previously reported but also exhibits two distinct structural phase transitions, facilitating sensing capabilities across two temperature ranges. Specifically, in the Li$_3$Sc$_2$(PO$_4$)$_3$ phase transitions from orthorhombic to monoclinic is observed around 550 K[21–27]. However, to the best of our knowledge another phase transition from monoclinic to orthorhombic phases at around 150 K is reported here for the first time These structural changes alter the point symmetry of RRR from EEEEE to QQQQ, and subsequently to WWW, thereby modulating the spectroscopic properties of $Eu^{3+}$ ions. In this work a comprehensive analysis of the temperature-dependent spectral and temporal responses of $Eu^{3+}$ ions in doped $Li_3Sc_2(PO_4)_3$ over an extensive temperature range is presented. Previous studies have demonstrated that dopant ion concentration can influence the phase transition temperature of materials; this effect has been corroborated in our work. Notably, for the first time, we demonstrate how variations in phosphor morphology can modulate the phase transition temperature and, consequently, the thermometric performance of both ratiometric and lifetime-based luminescence thermometers.

## 2. Experimental Section

*Materials and synthesis*

The powders of $Li_3Sc_2(PO_4)_3$ doped with $Eu^{3+}$ were synthesized by the solid-state reaction method using stoichiometric amounts of $Li_2CO_3$ (99.998% purity, Alfa Aesar), $Sc_2O_3$ (99.99% purity, Alfa Aesar), $NH_4H_2PO_4$ (99% purity, POL-AURA) and $Eu_2O_3$ (99.99% purity, Stanford Materials). Samples were ground in an agate mortar for 10 minutes. The hexane was used during the grinding process. Two series of $Li_3Sc_2(PO_4)_3$ samples were synthesized. The first group, doped with 1% $Eu^{3+}$, was annealed at temperatures varying from 1073 to 1573 K, while the second group of probes with varying $Eu^{3+}$ content was annealed at a constant temperature of 1373 K. All samples were prepared in alumina crucibles and annealed in air for 6 hours with a heating rate of 20 K min$^{-1}$. After annealing, the samples were ground in a mortar to prepare them for the measurements.

*Methods*

The X-ray diffraction (XRD) patterns were acquired using the PANalytical X'Pert Pro diffractometer with Cu Kα radiation (λ = 1.5418 Å). The measurement parameters were: V = 40 kV and I = 30 mA. The geometry used during examination was Bragg-Brentano. The characterization was performed over the 2θ range of 10° - 90°. The high-temperature powder diffraction studies were performed with the use of Anton Paar HTK 1200N in the temperature range 298-673 K every 25 K. The low-temperature studies were performed with the use of Oxford Cryosystems Phenix cryochamber in the temperature range 20-280 K every 20 K.

The differential scanning calorimetric (DSC) measurements were performed with the use of the Perkin-Elmer DSC 8000 calorimeter, equipped with Controlled Liquid Nitrogen Accessory LN2. A heating and cooling rate of measurements was 10 K min$^{-1}$. The powders were sealed in the aluminum pans.

The scanning electron microscope (SEM) FEI NovaNanoSEM 230 integrated with an energy-dispersive X-ray spectrometer (EDAX Apollo 40 SDD) with the resolution better than 135 eV and compatible with Genesis EDAX Microanalysis Software was used to acquire information about morphology and chemical composition of examined samples. Initially, the samples were positioned on the carbon stub to rectify issues pertaining to charging and drift effect. Subsequently, the samples were placed under the microscope and analysed using secondary (SE) and characteristic X-ray signals. The SEM images were recorded at an accelerating voltage of 5.0 kV in a beam deceleration mode in order to show more detailed features of the samples. In contrast, the EDS maps were performed at 30 kV from the selected grains of Eu$^{3+}$-doped Li$_3$Sc$_2$(PO$_4$)$_3$. Nevertheless, owing to the inherent limitations of the EDS method in terms of the detection of light elements (Z ≤ 4), the EDS map of the Li element was not included and discussed in the paper. The grain size distribution was determined using freeware software ImageJ (version 1.53n).

Raman spectra from 80 to 700 K were measured using a Renishaw inVia Basic Raman spectrometer equipped with a Leica confocal microscope DM2500, CCD detector, and laser operating at 514.5 nm. The temperature was controlled using a cryostat cell Linkam THMS600 equipped with quartz windows.

The photoluminescence emission spectra were obtained using the FLS1000 Fluorescence Spectrometer from Edinburgh Instruments, equipped with the 450 W Xenon lamp and the Hamamatsu R928P side window multiplier tube as a detector. The Xenon lamp was used as the excitation source. The luminescence decay curves were also measured using the FLS1000 Fluorescence Spectrometer equipped with the 150 W μFlash lamp. To control the temperature the THMS 600 heating-cooling stage from Linkam (0.1 K temperature stability and 0.1 K set point resolution) was used. The temperature range for both analyses was 83 K – 643 K. The temperature of the samples was stabilized for 2 minutes before measurement. The average lifetime ($\tau_{avr}$) of the excited states was determined by using of double – exponential function:

$$\tau_{avr} = \frac{A_1\tau_1^2 + A_2\tau_2^2}{A_1\tau_1 + A_2\tau_2} \qquad (1)$$

$$I(t) = I_0 + A_1 \cdot \exp\left(-\frac{t}{\tau_1}\right) + A_2 \cdot \exp\left(-\frac{t}{\tau_2}\right) \qquad (2)$$

where $\tau_1$ and $\tau_2$ are decay components and $A_1$ and $A_2$ are the amplitudes of the double – exponential function.

## 3. Results and discussion

In the monoclinic structure of $Li_3Sc_2(PO_4)_3$, two distinct types of $ScO_6$ octahedra are reported in the literature, although detailed analyses of their differences are rather scarce[21–28]. The variation between these octahedra arises primarily from the differing local environments of the *A* and *B* sites of $Sc^{3+}$ ions, which are influenced mainly by the surrounding $Li^+$ ions (Figure 1a). In $Li_3Sc_2(PO_4)_3$ structure, $Li^+$ cations are located in three inequivalent crystallographic positions. These differences in the local $Li^+$ environment result in variations in the $Sc^{3+}$-$O^{2-}$ bond lengths: for *A*, the bond lengths are relatively uniform, ranging from 2.047 to 2.180 Å. On the other hand, *B* exhibits a broader range of Sc-O bond lengths, from 1.990 to 2.213 Å, resulting in a more distorted octahedral geometry compared to the more regular $ScO_6$(*A*). When $Eu^{3+}$ ions are incorporated into the structure as dopants, they occupy both types of $Sc^{3+}$ sites. The presence of these two distinct crystallographic $Eu^{3+}$ environments is manifested in the emission spectra, which show contributions from both *A* and *B*, which will be discussed later in this work. For $Eu^{3+}$(*B*), the more distorted octahedral environment, characterized by lower local symmetry, results in greater splitting of the electronic energy levels. This leads to the observation of additional emission lines in the luminescence spectrum, reflecting the asymmetry of the *B* site occupied by $Eu^{3+}$.

$Li_3Sc_2(PO_4)_3$ can crystallize not only in a monoclinic structure but also in an orthorhombic form. The key structural difference between these two structures arises from the various distribution of lithium sites. In addition to differences in $Li^+$-$Li^+$ distances, these structures vary in the occupancy of $Li^+$ sites. In the orthorhombic structure, all $Li^+$ sites are partially occupied, whereas in the monoclinic phase, $Li^+$ sites are either fully occupied or completely unoccupied. From a luminescence perspective, the most crucial consequence lies in the local environment of $Sc^{3+}$ cations, which are replaced by the $Eu^{3+}$ dopant. In the monoclinic structure, as previously mentioned, two types of scandium coordination exist. In contrast, the orthorhombic phase features a single type of $ScO_6$ octahedra, with $Sc^{3+}$-$O^{2-}$ bond lengths ranging from 1.994 Å to 2.162 Å. The analysis of the room temperature XRD patterns of the $Li_3Sc_2(PO_4)_3$:1%$Eu^{3+}$ annealed at different temperatures revealed that crystallization of the pure phase of $Li_3Sc_2(PO_4)_3$ corresponding to the monoclinic structure (ICSD 50420) is observed at temperature above 1173K (Figure 1b). Further increase in the annealing temperature leads to the slight narrowing of the reflections due to the change in the morphology of the obtained materials as it will be discussed in the later part of this work. Below 1173 K additional reflections can be found in the XRD pattern.

The phase in which $Li_3Sc_2(PO_4)_3$ crystallizes depends on various factors, including synthesis conditions and compound composition. Notably, the incubation temperature can also influence its structure, leading to a first-order structural phase transition. To date, two reversible, temperature-induced phase transitions have been reported in the literature: from α to β (both monoclinic structures) around 460 K, and from β to orthorhombic $\gamma_{HT}$-structure around 520 K. Interestingly, to the best of our knowledge, no structural studies of $Li_3Sc_2(PO_4)_3$ have been conducted at lower than room-temperature, likely due to its primary application as an ionic conductor for battery technologies. To investigate the influence of the temperature on the structure of the $Li_3Sc_2(PO_4)_3$:$Eu^{3+}$ the thermal-dependent in situ XRD patterns were measured (Figure S1-S30) and based on the Rietveld refinement of the obtained patterns the contribution of particular crystallographic phases of $Li_3Sc_2(PO_4)_3$ in the volume of the sample was determined (Figure 1c). At 20 K the XRD pattern of $Li_3Sc_2(PO_4)_3$:1%$Eu^{3+}$ corresponds to the $\gamma_{LT}$. However an increase in temperature above 100 K results in the appearance of the reflection corresponding to the monoclinic α phase, which starts to dominate above 160 K and persist in the 160-500 K temperature range. Above 500 K additional reflections corresponding to the orthorhombic $\gamma_{HT}$ phase of $Li_3Sc_2(PO_4)_3$ start to be observed and around 520 K this phase becomes the only one observed in the XRD pattern. The difference between α and β phases of the $Li_3Sc_2(PO_4)_3$ cannot be recognized based on the performed crystallographic analysis. Therefore in the further part of the text the monoclinic phase of the $Li_3Sc_2(PO_4)_3$ will be denoted as α/β. Obtained results confirmed the thermally induced changes in the structure of the $Li_3Sc_2(PO_4)_3$. Based on the DSC measurements it was found that the $Eu^{3+}$ concentration affects temperature of both $\gamma_{LT}\rightarrow\alpha/\beta$ and $\alpha/\beta\rightarrow\gamma$ phase transitions. The $\gamma_{LT}\rightarrow\alpha/\beta$ increases from 164 K for 0.1%$Eu^{3+}$ up to 181 K for 2%$Eu^{3+}$. However the opposite trend is observed for $\alpha/\beta\rightarrow\gamma$ where $T_{PT}$ decreases from 530 K for 0.1% $Eu^{3+}$ to 508 K for 2% $Eu^{3+}$ (Figure 1d). Observed concentration effect is well known in the literature and results from the growing concentration of the dopant ions of the different ionic radii in respect to the host material cations[15,16,19,29]. The morphological studies indicate that the $Li_3Sc_2(PO_4)_3$:$Eu^{3+}$ consists of aggregated crystals of the average particle size that depends on the annealing temperature as it will be discussed later in this paper. However in the case of the $Li_3Sc_2(PO_4)_3$:$Eu^{3+}$ annealed at 1173 K crystals of around 4.8 μm in diameter (Figure 1e, see also Figure S31-S35). Noteworthy, the uniform elements distribution was confirmed by the EDS studies (Figure 1f-h).

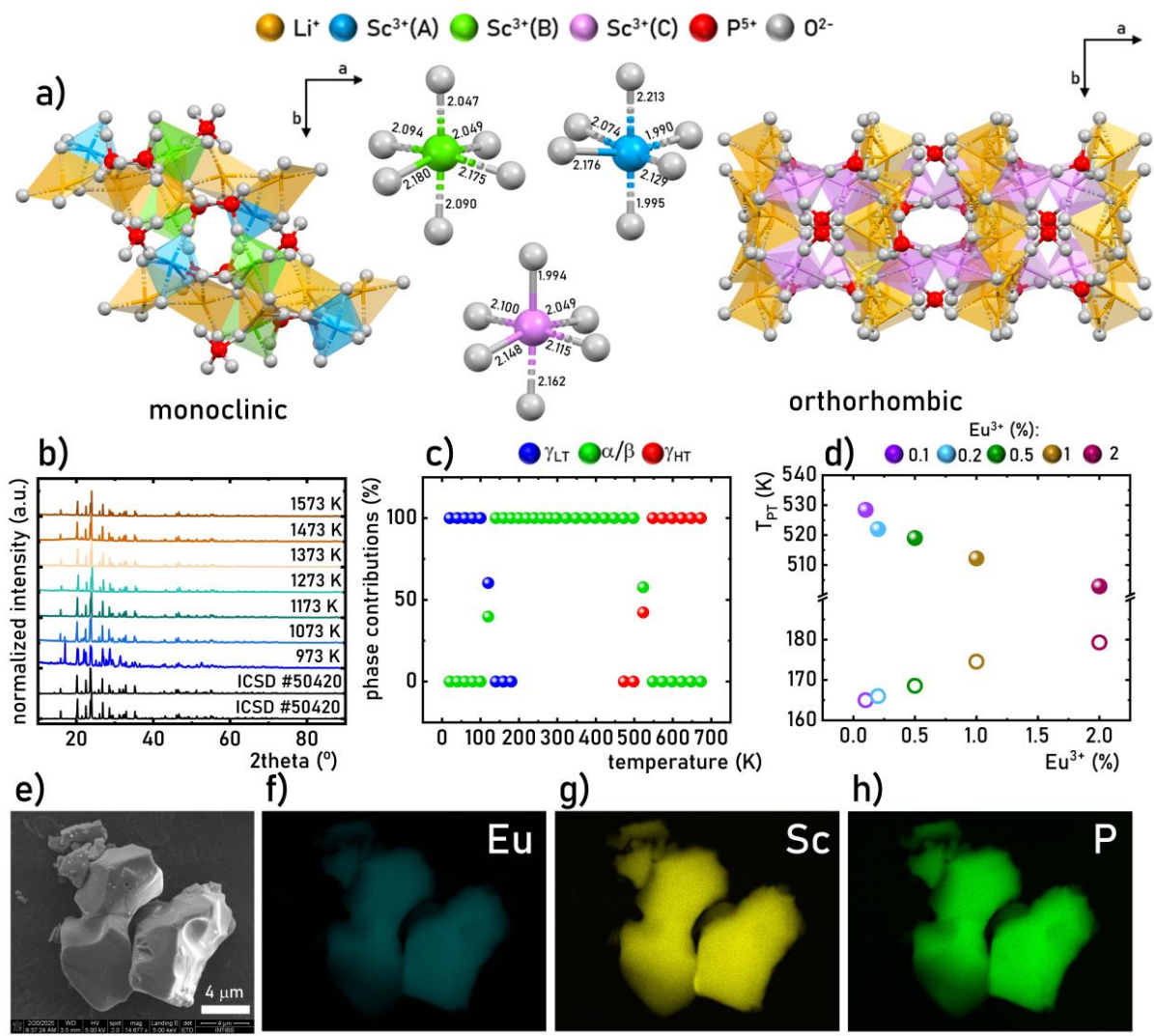

**Figure 1.** The visualization of the monoclinic and orthorhombic structures of $Li_3Sc_2(PO_4)_3$ -a); the comparison of the room temperature XRD patterns for $Li_3Sc_2(PO_4)_3$:1%$Eu^{3+}$ annealed at different temperatures -b); the contribution of different crystallographic phases of $Li_3Sc_2(PO_4)_3$ as a function of temperature determined from the analysis of XRD patterns-c); the influence of the $Eu^{3+}$ concentration on the $\gamma_{LT} \rightarrow \alpha/\beta$ (open circles) and $\alpha/\beta \rightarrow \gamma_{HT}$ (full circles) phase transition temperatures -d); the representative SEM image of $Li_3Sc_2(PO_4)_3$:1%$Eu^{3+}$ - e) and elemental distribution of Eu -f); Sc-g) and P-h).

To get more information on phase transitions, the Raman spectra were measured for $Li_3Sc_2(PO_4)_3$:0.1% $Eu^{3+}$ as a function of temperature (Figure 2a). All spectra exhibit typical bands observed for orthophosphates; the tentative assignment for selected temperatures, based on literature data and similar compounds, is proposed in Table S1 [30–34].

Based on crystal structures, the factor group analysis predicts the same number of Raman-active phonon modes for α/β and $\gamma_{HT}$ phases; therefore, during the α/β → $\gamma_{HT}$ phase transition with increasing symmetry, merging of bands is expected only due to thermal broadening and overlapping. At 80 K, all Raman bands are narrow, indicating that the crystal structure is fully ordered. The spectrum changes substantially at 160 K, demonstrating various distortions of orthophosphate ions in the monoclinic phase. A coexistence of both the α and $\gamma_{LT}$ phases at 150 K confirms the discontinuous nature of $\gamma_{LT}$ → α transformation (Figure 2b). Near 485 K, where the isosymmetric α → β transformation is expected[35], no phonon softening or hardening was registered, indicating that this isosymmetric phase transition is accompanied by small changes in structural parameters. Figure 2c demonstrates the changes in full width at half maximum (fwhm) for the band having a strong contribution of the $\nu LiO_4$ vibrational mode (about 515 cm$^{-1}$). This peak broadens strongly upon heating, and the fwhm increase becomes faster above around 480 K, suggesting that the Li$^+$ ions start disordering. This finding is consistent with previous Raman studies on isostructural $Li_3Fe_2(PO_4)$, exhibiting α → β similar phase transition at 485 K, caused by released Li$^+$ occupancy[30], and electrical studies on $Li_3Sc_2(PO_4)_3$ that showed a strong increase in conductivity in monoclinic β phase[35].

Continuous heating leads to a strong broadening of bands above 540 K, evidencing a second phase transition to the orthorhombic polymorph $\gamma_{HT}$. In this phase, the fwhm of the $\nu LiO_4$ band becomes so high that the peak cannot be fitted anymore, indicating stronger occupational disorder. The $\nu_{as}PO_4$ is less affected up to 530 K, until the 6.8 cm$^{-1}$ jump of fwhm occurs during the α/β → $\gamma_{HT}$ phase transition. The Raman spectrum of the $\gamma_{HT}$ polymorph exhibits only a few bands and resembles the spectrum obtained at 4.5 GPa for the high-pressure phase obtained for $Li_3Sc_2(PO_4)_3$:1% $Cr^{3+}$ crystals[36], see Figure 2a. This result suggests that the high-pressure phase ($\gamma_{HP}$) is also orthorhombic and is at least partially disordered, which allows Li$^+$ ions mobility.

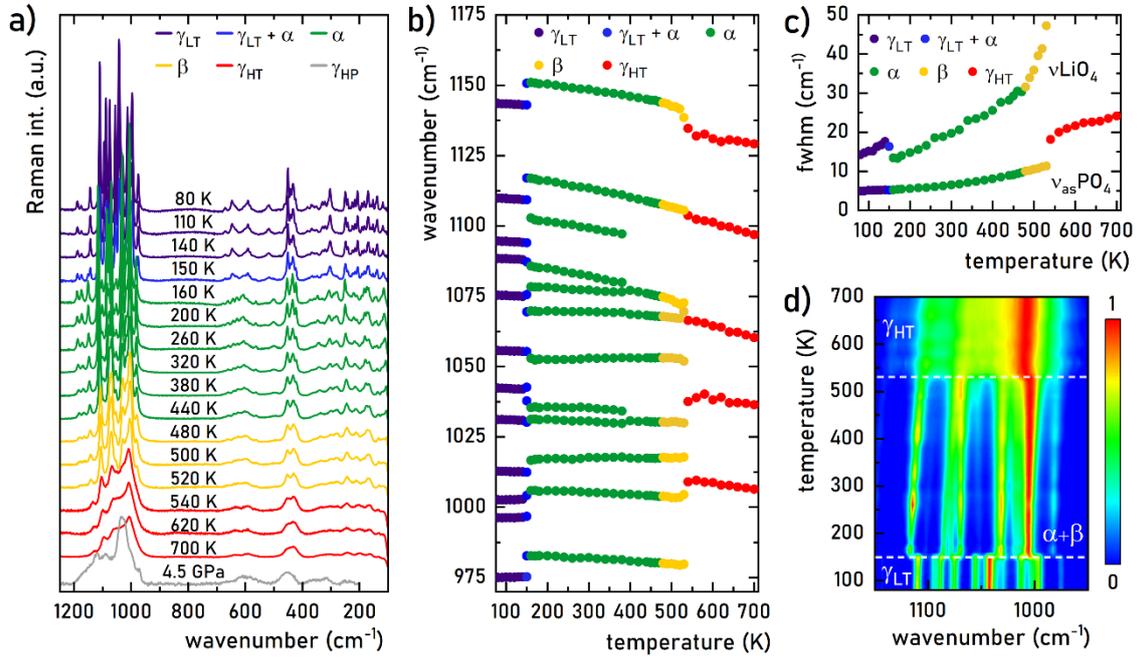

**Figure 2.** The thermal evolution of Raman spectra of Li$_3$Sc$_2$(PO$_4$)$_3$:0.1% Eu$^{3+}$ compared to a high-pressure Raman spectrum measured at about 4.5 GPa[36] -a); the change in selected band positions as a function of temperature-b); changes in fwhm for phonon modes at about 1110 cm$^{-1}$ ($\nu_{as}$PO$_4$) and 515 cm$^{-1}$ ($\nu$LiO$_4$)-c); Raman normalized intensity map in the 1150-950 cm$^{-1}$ spectral range-d).

The luminescent properties of phosphors doped with Eu$^{3+}$ ions are primarily determined by the interconfigurational *4f-4f* transitions associated with the radiative depopulation of the $^5D_0$ state. Spin-orbit coupling causes the splitting of the $^7F$ level into multiple energy states, labelled $^7F_0$ to $^7F_6$. Electronic transitions between the excited $^5D_0$ excited state and the $^7F_J$ levels result in the formation of characteristic emission bands at approximately 580 nm, 590 nm, 620 nm, 650 nm, and 730 nm, corresponding to the $^5D_0\rightarrow{}^7F_J$ transitions ($^5D_0\rightarrow{}^7F_0$, $^5D_0\rightarrow{}^7F_1$, $^5D_0\rightarrow{}^7F_2$, $^5D_0\rightarrow{}^7F_3$, $^5D_0\rightarrow{}^7F_4$, respectively) (Figure 3a). These emission bands are consistently observed in all Li$_3$Sc$_2$(PO$_4$)$_3$:Eu$^{3+}$ phases (Figure 3b). Notably, the $^5D_0\rightarrow{}^7F_0$ transition is typically only observed in the host materials of very specific symmetry e.g., $C_s$, $C_v$, $C_3$. The presence of this transition in all $Li_3Sc_2(PO_4)_3:Eu^{3+}$ phases confirms the low-symmetry nature of these structures. Each $^7F_J$ level undergoes Stark splitting due to interactions with the electric field of the host material, and the number of Stark components depends on both the $J$ quantum number and the point symmetry of the crystallographic site occupied by $Eu^{3+}$ ions. A general trend is observed: higher $J$ values and lower symmetry correspond to an increased number of Stark sublevels. Importantly, neither the $^5D_0$ excited level nor the $^7F_0$ ground level splits under the Stark effect. Therefore, for each crystallographically distinct site occupied by $Eu^{3+}$ ions, this transition appears as a single emission line in the spectrum. Detailed analysis of the $^5D_0 \rightarrow ^7F_0$ band reveals that in the $Li_3Sc_2(PO_4)_3:Eu^{3+}$ phase, two distinct crystallographic positions are occupied by $Eu^{3+}$ ions, which correspond to two inequivalent $Sc^{3+}$ sites. This finding will be further explored in the paper. Additionally, the low symmetry of $Li_3Sc_2(PO_4)_3:Eu^{3+}$ is corroborated by the relative intensity of the $^5D_0 \rightarrow ^7F_2$ electric-dipole transition compared to the $^5D_0 \rightarrow ^7F_1$ magnetic-dipole transition, whose intensity is independent of the host material[37–39]. The luminescence intensity ratio ($LIR_1$) of these intensities, serves as an effective indicator of the local symmetry surrounding $Eu^{3+}$ ions and can used to analyze structural phase transitions in host materials:

$$LIR_1 = \frac{\int_{610nm}^{630nm} \left(^5D_0 \rightarrow ^7F_2\right) d\lambda}{\int_{580nm}^{600nm} \left(^5D_0 \rightarrow ^7F_1\right) d\lambda} \qquad (3)$$

Luminescence spectra of $Li_3Sc_2(PO_4)_3:1\%Eu^{3+}$ measured at 83 K, 293 K, and 550 K which can be considered as a representative for $\gamma_{LT}$, $\beta$ and $\gamma_{HT}$ phases indicate that $LIR_1$ assumes values of 1.65, 1.95, and 1.8 for these structures, respectively. To facilitate interpretation of the changes in luminescence spectra corresponding to different structural phases, normalized intensity maps are presented for the different crystallographic phases of $Li_3Sc_2(PO_4)_3:1\%Eu^{3+}$.

The analysis focuses on the $^5D_0 \to {}^7F_1$ (Figure 3c) and $^5D_0 \to {}^7F_2$ (Figure 3d) bands, as higher $^7F_J$ levels exhibit increased Stark splitting and spectral overlap of the Stark lines, complicating precise analysis. Phase transitions primarily influence the energy of the Stark levels, which manifest as shifts in the spectral positions of emission peaks. These transitions also alter the relative intensity ratios of the $^5D_0 \to {}^7F_1$ and $^5D_0 \to {}^7F_2$ bands. A reduction in the number of Stark lines observed in spectra recorded for the $\gamma_{HT}$ phase may be attributed to spectral broadening caused by thermal effects, which increase spectral overlap. Nevertheless, shifts in the emission line positions provide unambiguous evidence for structural phase transitions. The structural phase transitions $\gamma_{HT} \to \alpha/\beta$ result in an initial increase in $LIR_1$, followed by a slight decrease for $\alpha/\beta \to \gamma$, suggesting a temporary reduction followed by increase in symmetry during the phase transition (Figure 3e). In the case of the $Eu^{3+}$ luminescence the single exponential decay is expected. However in the temporal range where the structural phase transition of the $Li_3Sc_2(PO_4)_3$:$Eu^{3+}$ is expected the coexistence of signals from different phases of the phosphor hinders the fitting of the luminescence decay curve using single exponential curve. Therefore, the average lifetime ($\tau_{avr}$) was determined according to the procedure given in the Experimental section. Interestingly, the $\tau_{avr}$ of the $^5D_0$ level exhibits an inverse trend in respect to the $LIR_1$, with initial $\tau_{avr}$ values of 2.5 ms for the $\gamma_{LT}$ phase decreasing to 2.25 ms for the $\alpha/\beta$ phase before increasing to 2.5 ms in the $\gamma$ phase (Figure 3e). Given that the $^5D_0$ level is separated from the $^7F_6$ level by approximately 12,000 cm$^{-1}$, the probability of nonradiative depopulation of the excited state through multiphonon processes is minimal. Therefore, the observed differences in $\tau_{avr}$ across the structural phases reflect variations in the radiative depopulation probability of the $^5D_0$ level. Although interionic interactions are not expected to significantly influence luminescence in $Eu^{3+}$-doped phosphors, prior studies indicate that altering the dopant ion concentration can affect the phase transition temperature and thereby modify the contributions of individual phases to the luminescence spectrum.

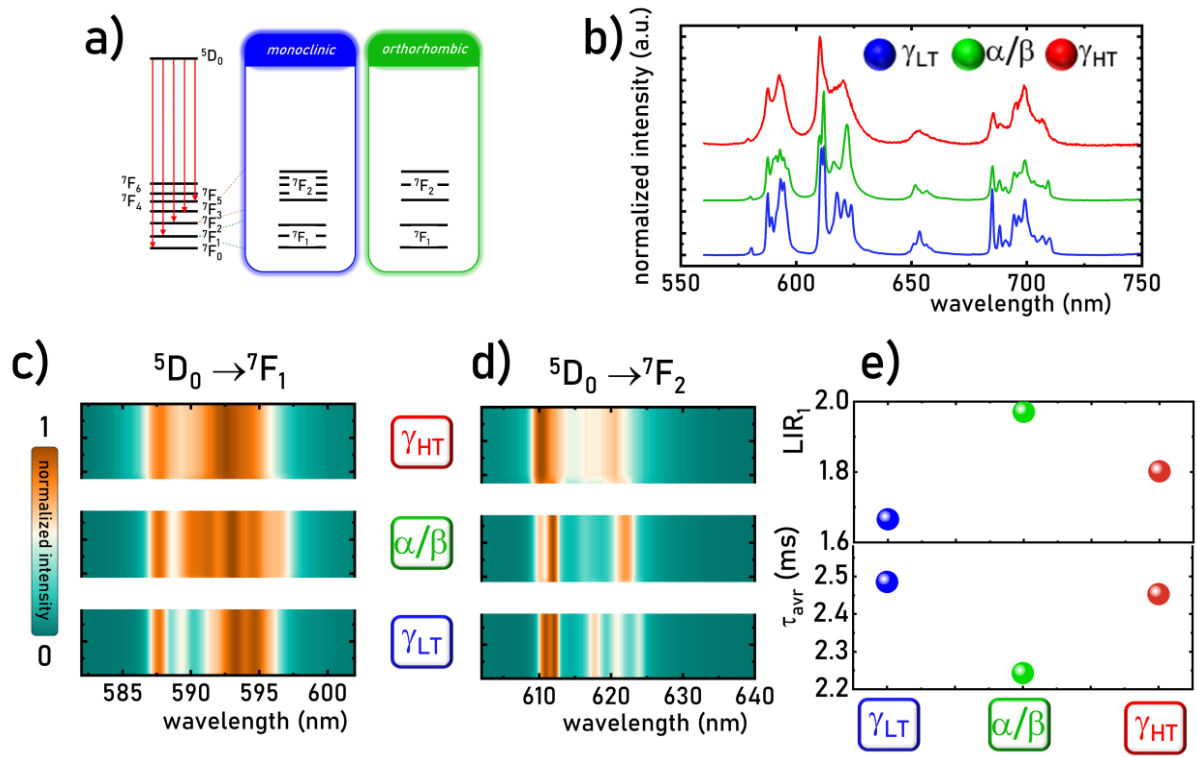

**Figure 3.** Simplified energy diagram of $Eu^{3+}$ ions in the $Li_3Sc_2(PO_4)_3$:$Eu^{3+}$ with the splitting of the $^7F_1$ and $^7F_2$ multiplets in different symmetries of the $Li_3Sc_2(PO_4)_3$-a); emission spectra of $Li_3Sc_2(PO_4)_3$:0.1%$Eu^{3+}$ measured at 83 K (blue line), 263 K (green line) and 623 K (red line) which can be treated as a representative for low temperature 83 K (LT); medium temperature 293 K (MT) and high temperature 563 K (HT) phases of the $Li_3Sc_2(PO_4)_3$:$Eu^{3+}$ -b); luminescence intensity maps of the $Li_3Sc_2(PO_4)_3$:1%$Eu^{3+}$ for different phases of the host material presented in the spectra ranges corresponding to the $^5D_0\rightarrow{}^7F_1$ -c) and $^5D_0\rightarrow{}^7F_2$ -d) electronic transitions; the influence of the phase of the $Li_3Sc_2(PO_4)_3$:$Eu^{3+}$ on the $\tau_{avr}$ of the $^5D_0$ state and the *LIR$_1$* -e).

As it was already introduced in the $Li_3Sc_2(PO_4)_3$:$Eu^{3+}$ compound, $Sc^{3+}$ ions occupy two distinct crystallographic sites, both of which can be substituted by $Eu^{3+}$ ions. Due to the differing local environments of these sites, variations in the optical responses of $Eu^{3+}$ ions in these positions are anticipated. Experimental observations confirm this, as the emission spectra of $Li_3Sc_2(PO_4)_3$:$Eu^{3+}$ reveal contributions from both sites (Figure 4a-c, Figure S36, S37). When excited at a $\lambda_{exc}$=392 nm, luminescence originates predominantly from site *A*. Adjusting the excitation wavelength to $\lambda_{exc}$=397 nm enables simultaneous excitation of $Eu^{3+}$ ions at both sites. Notably, each crystallographic phase of $Li_3Sc_2(PO_4)_3$:$Eu^{3+}$ exhibits distinct emission spectra. Given that the spectral positions of the *4f–4f* electronic transitions of $Eu^{3+}$ are minimally influenced by the host lattice composition, emissions from sites *A* and *B* overlap spectrally. Consequently, even at 83 K, isolating luminescence exclusively from site *B* remains challenging. The most pronounced differences of the optical response of $Eu^{3+}$ ions ad *A* and *B* sites are observed in the $^5D_0 \rightarrow {}^7F_4$ transition, where Stark splitting leads to shifts in spectral lines. The $^5D_0 \rightarrow {}^7F_0$ transition, typically represented by a single line for each site of $Eu^{3+}$ ions, serves as a clear indicator of $Eu^{3+}$ occupancy at both sites (Figure 4d). Under $\lambda_{exc}$=392 nm and $\lambda_{exc}$=394 nm excitation, a single emission line appears around 580.5 nm. However, excitation at $\lambda_{exc}$=397 nm reveals an additional component near 579.5 nm. Considering that stronger interactions between $Eu^{3+}$ and surrounding $O^{2-}$ ions, associated with shorter $Eu^{3+}$- $O^{2-}$ bond lengths, can elevate the energy of the emitting level, it is plausible to assign the 579.5 nm line to site *B* and the 580.5 nm line to site *A*. The presence of a second component in the $^5D_0 \rightarrow {}^7F_0$ emission band upon $\lambda_{exc}$=397 nm excitation correlates with an increased number and intensity of Stark components in the 690 - 705 nm range of the $^5D_0 \rightarrow {}^7F_4$ transition (Figure 4e). Given that this emission range overlaps with that of $Cr^{3+}$ ions, excitation spectra were recorded at $\lambda_{em}$=699 nm to investigate this possibility. The absence of characteristic broad absorption bands of $Cr^{3+}$ ions in these spectra effectively rules out the presence of $Cr^{3+}$ impurities in the material. Comparative analysis of excitation spectra for both crystallographic sites reveals slight spectral shifts in absorption bands (Figure 4f). Temperature-dependent emission studies indicate that the *$LIR_1$* parameter is higher for site *B* in respect to site *A*, suggesting a lower local symmetry around $Eu^{3+}$ ions at site *B*. Both sites exhibit a sharp increase in *$LIR_1$* at the $\gamma_{LT} \rightarrow \alpha/\beta$ phase transition temperature, followed by a decrease during the $\alpha/\beta \rightarrow \gamma$ transition (Figure 4g). Additionally, $Eu^{3+}$ ions at site *A* demonstrate slightly greater thermal stability compared to those at site *B* (Figure 4h). Furthermore, analysis of the temperature dependence of the $^5D_0$ level luminescence decay reveals that the $\tau_{avr}$ at 83 K is 2.6 ms for site *A* and 1.7 ms for site *B*, indicating a lower probability of radiative processes for $Eu^{3+}$ ions at site *A* (Figure 4i).

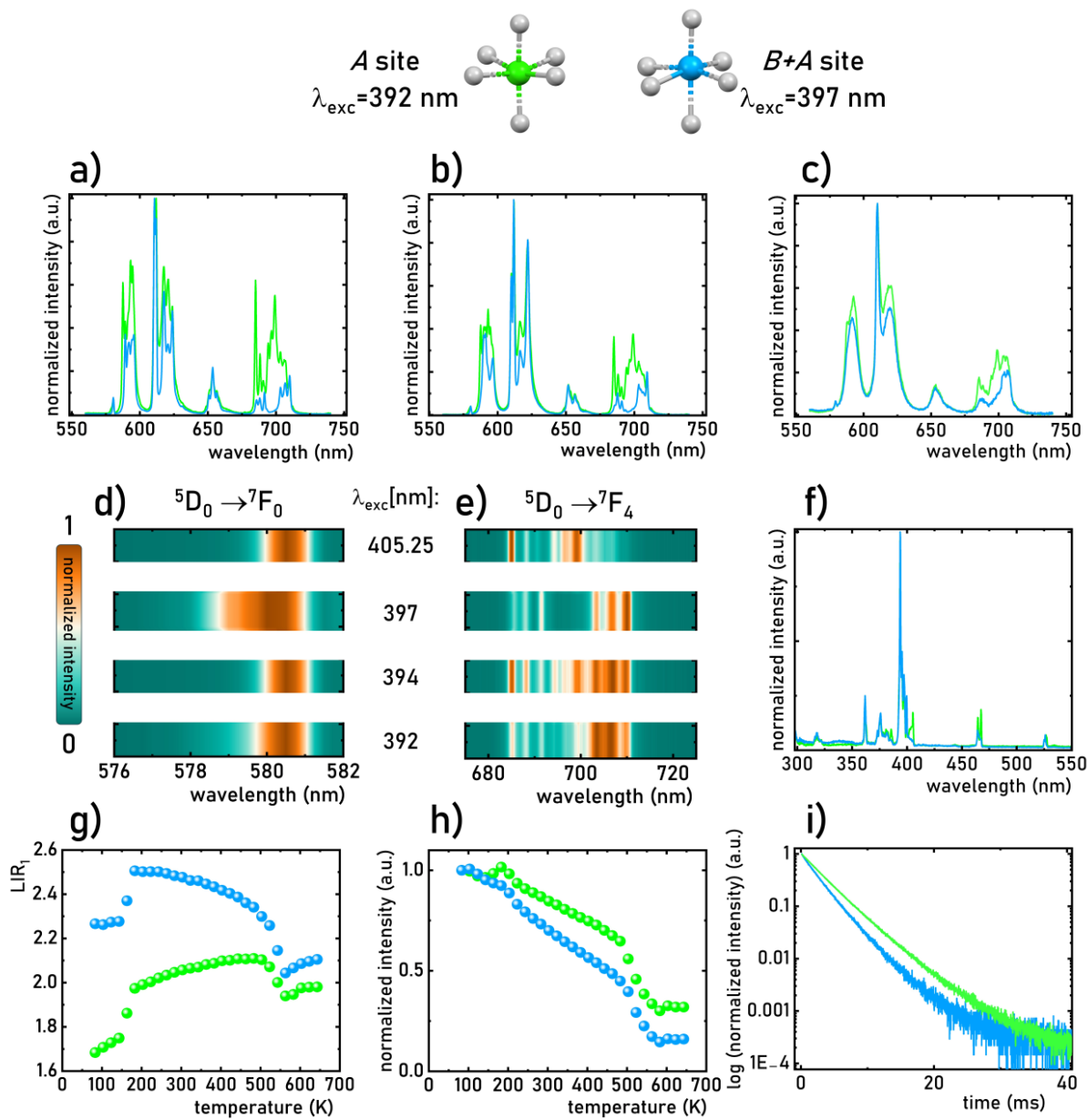

**Figure 4**. Comparison of the emission spectra of $Eu^{3+}$ upon $\lambda_{exc}$=392 nm (green line) and $\lambda_{exc}$=397 nm (blue line) of $Li_3Sc_2(PO_4)_3$:$Eu^{3+}$ measured at 83K -a); 300K-b) and 603 K-c); luminescence maps of $Li_3Sc_2(PO_4)_3$:$Eu^{3+}$ obtained upon different excitation wavelengths limited to the spectral range corresponding to the $^5D_0\rightarrow{}^7F_0$ -d) and the $^5D_0\rightarrow{}^7F_4$ transitions -d); comparison of the excitation spectra of $Li_3Sc_2(PO_4)_3$:$Eu^{3+}$ measured for $\lambda_{em}$=703

nm (green line) and $\lambda_{em}$=695 nm (blue line)-f); thermal dependence of $LIR_1$-g); integral emission intensity -h) and luminescence decay profiles -i) for $Eu^{3+}$ ions in *A* (green dots) and *B* (blue dots) sites.

The concentration of dopant ions significantly influences the phase transition temperature in materials due to the disparity in ionic radii between the dopant and the host cations[16,18,29]. To investigate this effect, the spectroscopic properties of $Li_3Sc_2(PO_4)_3$:$Eu^{3+}$ were examined across varying $Eu^{3+}$ ion concentrations (Figure 5a). At low dopant levels, emission spectra at room temperature predominantly display luminescence associated with the *A* site. As the $Eu^{3+}$ ions concentration increases, there is a notable rise in the emission intensity associated with the *B* site, especially evident in the $^5D_0 \rightarrow {}^7F_4$ emission band. Additionally, an increase in dopant concentration leads to a decrease in the intensity ratio of the $^5D_0 \rightarrow {}^7F_2$ to $^5D_0 \rightarrow {}^7F_1$ transitions, further indicating enhanced *B* site luminescence contribution. The $LIR_1$ decreases monotonically with higher $Eu^{3+}$ ion concentrations, directly suggesting increased occupancy of the *B* site (Figure 5b). Considering that emission in the 699 - 700 nm spectral range is attributed to $Eu^{3+}$ ions at the *B* site, while this in the 682-683 nm range correspond to the *A* site, the intensity ratio of these spectral regions may serve as an indirect measure of relative $Eu^{3+}$ sites occupancy (Figure 5c). Despite potential spectral overlaps, this ratio reveals that elevating $Eu^{3+}$ ion concentration from 0.1% to 2% results in a twelvefold increase in *B* site luminescence in respect to the *A* site, indicating significant *B* site occupancy at higher dopant levels. Notably, up to 1% $Eu^{3+}$ concentration, this ratio only doubles, suggesting a preferential occupancy of the *A* site by $Eu^{3+}$ ions for low dopant concentration, with substantial *B* site occupation occurring at higher concentrations. Analysis of luminescence decay kinetics supports these findings (Figure 5d). For $Eu^{3+}$ ion concentrations below 1%, the $\tau_{avr}$ remains around 2.25 ms. However, at 5% $Eu^{3+}$, $\tau_{avr}$ shortens to 2.05 ms. The fact that $\tau_{avr}$ does not reach the value characteristic of the *B* site even at 2% $Eu^{3+}$ is likely due to spectral overlap between emissions from both crystallographic positions of $Sc^{3+}$ ions. These observations underscore the intricate relationship between dopant concentration and site occupancy, which in turn affects the material's luminescent properties and phase transition behavior.

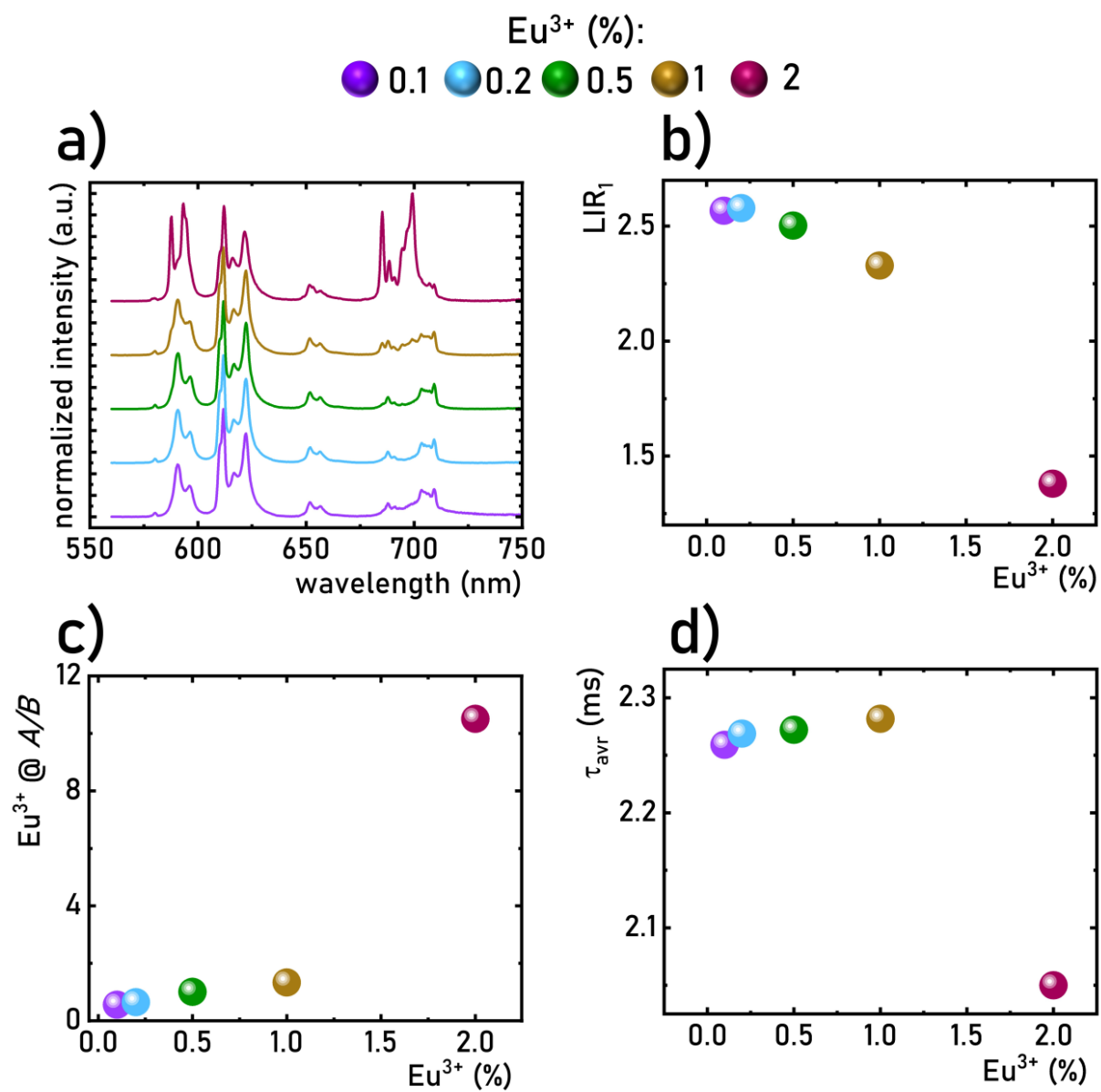

**Figure 5**. The influence of the $Eu^{3+}$ ions concentration on the room temperature emission spectra of $Li_3Sc_2(PO_4)_3$:$Eu^{3+}$-a); the influence of $Eu^{3+}$ ions concentration on the room temperature $LIR_1$-b); influence of the $Eu^{3+}$ ions concentration on the $Sc^{3+}$ sites occupation in $Li_3Sc_2(PO_4)_3$:$Eu^{3+}$-c) and the $\tau_{avr}$-d).

The observed variations in the Eu$^{3+}$ emission spectrum with increasing temperature, attributed to structural phase transitions in the Li$_3$Sc$_2$(PO$_4$)$_3$:Eu$^{3+}$, offer opportunities for temperature sensing applications (Figures S38-S42). While the thermal evolution of the *LIR$_1$* indicates the occurrence of these phase transitions, the rate of change in *LIR$_1$* is insufficient for precise temperature measurements. This limitation arises from the spectral overlap of Eu$^{3+}$ emission bands corresponding to each phase. Consequently, a more effective approach involves analyzing the thermal intensity variations of Stark lines associated with individual Eu$^{3+}$ phases. Spectral analysis suggests that maximizing the thermal response of *LIR* changes can be achieved by selecting the following spectral ranges for analysis:

$$LIR_2 = \frac{\int_{621nm}^{624nm} \left(^5D_0 \to {}^7F_2\right) d\lambda}{\int_{652nm}^{655nm} \left(^5D_0 \to {}^7F_3\right) d\lambda} \qquad (4)$$

Consistent thermal behavior of *LIR$_2$* is observed across all Eu$^{3+}$ concentrations: as temperature increases from 83 K, *LIR$_2$* gradually rises, with a pronounced increase around 150 K corresponding to the $\gamma_{LT} \to \alpha$ phase transition (Figure 6a). Further temperature elevation leads to a continuous decrease in *LIR$_2$* until approximately 550 K, where its rapid decrease indicates the $\beta \to \gamma$ phase transition. For accurate temperature readings using a luminescent thermometer, a monotonic *LIR* change within the analyzed thermal range is essential. Thus, two operational thermal ranges for the Li$_3$Sc$_2$(PO$_4$)$_3$:Eu$^{3+}$ can be identified (marked in blue and green in Figure 6a). To quantify the thermal variations of *LIR$_2$* and assess its potential for remote temperature sensing, the relative thermal sensitivity (*S$_R$*) is calculated as follows:

$$S_R = \frac{1}{LIR} \frac{\Delta LIR}{\Delta T} \cdot 100\% \qquad (5)$$

where $\Delta LIR$ represents the change of $LIR$ corresponding to the $\Delta T$ change of temperature. In the first thermal range, encompassing the $\gamma_{LT}\rightarrow\alpha$ phase transition, maximum $S_R$ values of 7.8% K$^{-1}$ are achieved for 0.1% Eu$^{3+}$ at 148 K (Figure 6b). Increasing the Eu$^{3+}$ concentration results in a gradual decrease in $S_{Rmax}$ and an upward shift in the temperature at which $S_{Rmax}$ occurs. A detailed analysis reveals a linear decrease in $S_{Rmax}$ (Figure 6c) and a linear increase in $T@S_{Rmax}$ (Figure 6d) with rising dopant concentration. Despite the high $S_R$ values, the narrow operational thermal range characteristic of phase-transition-based luminescent thermometers poses a limitation[16,20,40]. However, adjusting the dopant ion concentration can shift the thermometer's operational range, optimizing its performance for specific applications[16,19]. The second thermal range is broader, but the $S_R$ values are significantly lower (Figure 6e). The maximum $S_R$ of approximately 0.65% K$^{-1}$ is observed at 520 K for 0.5% Eu$^{3+}$, corresponding to the $\beta\rightarrow\gamma$ phase transition. Unlike the first thermal range, there is no direct correlation between both $S_{Rmax}$ and Eu$^{3+}$ ions concentration (Figure 6f). However the $T@S_{Rmax}$ monotonically decreases with dopant concentration in agreement with the change in the phase transition temperature (Figure 6g). Although a monotonic change in $LIR_2$ is present throughout this thermal range, the low luminescence intensity of the γ phase and minimal thermal dynamics associated with α-phase emission result in $S_R$ values not exceeding 0.2% K$^{-1}$ up to about 450 K-a behavior anticipated for Eu$^{3+}$ ions. While $LIR_2$ spikes are also observed during the $\beta\rightarrow\alpha$ transition, the lower $S_{Rmax}$ values are due to thermal broadening of Stark lines, leading to spectral overlap of signals from both phases and reduced thermal dynamics of $LIR_2$ changes. In summary, the optimal operational ranges for the ratiometric luminescent thermometer based on Li$_3$Sc$_2$(PO$_4$)$_3$:Eu$^{3+}$ are 120 -190 K and 480 - 600 K, corresponding to the structural phase transitions in the host material. Despite the opposite monotonicity of $LIR_2$ changes in these two thermal ranges, their sufficient thermal separation ensures reliable temperature measurements. Although these operational ranges may appear narrow, they are adequate for various applications, and the high relative sensitivity enhances the quality of temperature readings. Comparatively, this material exhibits the highest reported sensitivity values below 200 K among all phase-transition-based thermometers and uniquely offers two operational thermal ranges. However, disregarding the operational ranges, LiYO$_2$:Eu$^{3+}$ demonstrates higher $S_R$ values of up to 12.4% K$^{-1}$[17].

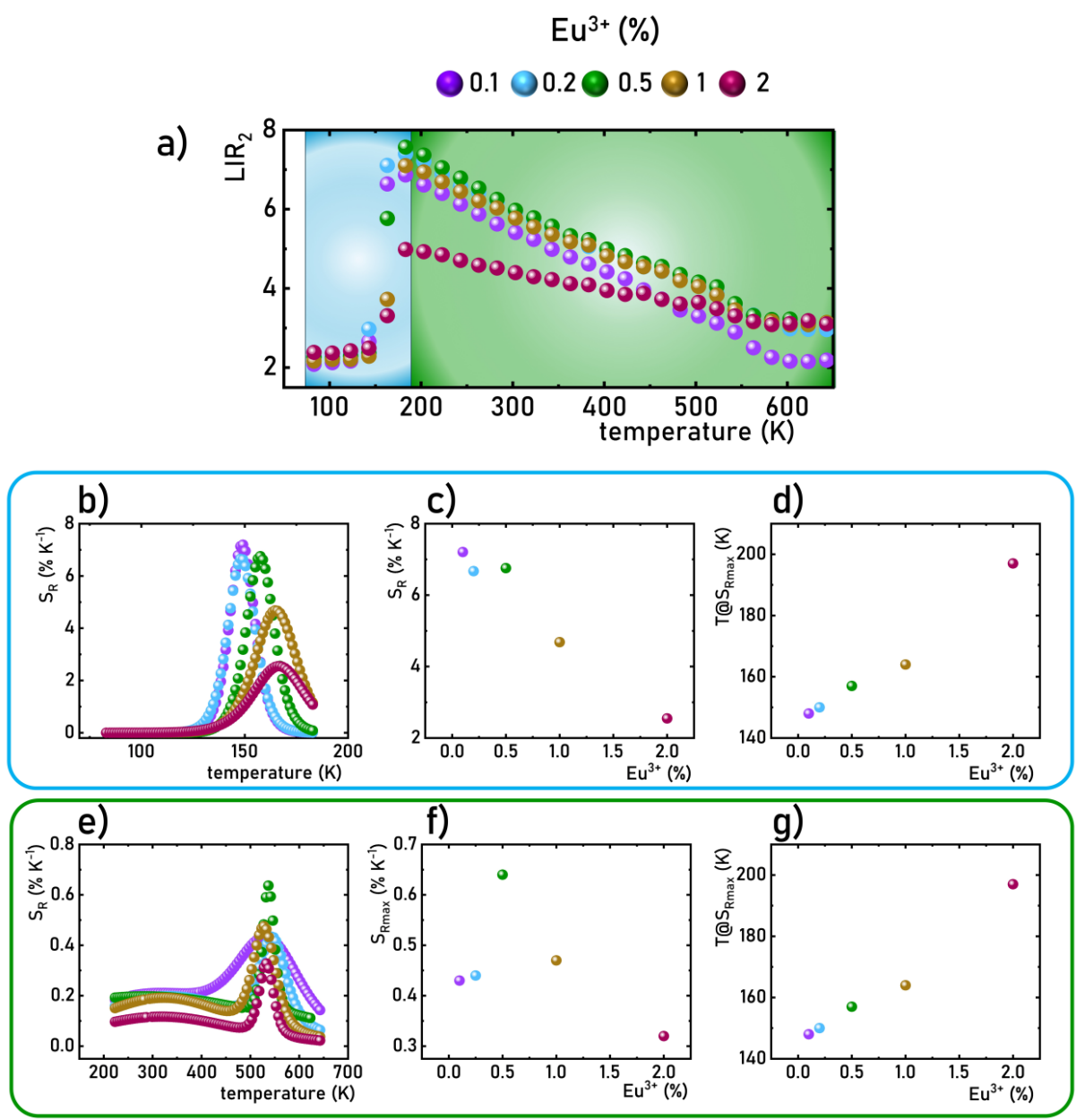

**Figure 6.** Thermal dependence of $LIR_2$ for different concentrations of Eu$^{3+}$ ions with denoted two thermal operating ranges (Range 1 in blue and Range 2 in green) of luminescence thermometers-a); thermometric performance of the ratiometric luminescence thermometer operating in the Range 1: thermal dependence of $S_R$-b) the influence of the Eu$^{3+}$ dopant concentration on the $S_{Rmax}$-c) and $T@S_{Rmax}$-d); thermometric performance of the ratiometric luminescence thermometer operating in the Range 2: thermal dependence of $S_R$-e) the influence of the $Eu^{3+}$ dopant concentration on the $S_{Rmax}$-f) and $T@S_{Rmax}$-g).

Not only the shape of the $Eu^{3+}$ emission spectrum depend on the local symmetry of the crystallographic site, but also the luminescence kinetics of the $^5D_0$ level. To investigate this effect in $Li_3Sc_2(PO_4)_3$:$Eu^{3+}$, luminescence decay profiles were recorded as a function of temperature for various $Eu^{3+}$ ion concentrations (Figure 7a, S43-S47). For dopant levels up to 1% $Eu^{3+}$, the $\tau_{avr}$ initially reached approximately 2.5 ms at 83 K and decreases to about 2.3 ms above the $\gamma_{HT} \rightarrow \alpha$ phase transition temperature, and then gradually increases with rising temperature, eventually approaching the initial value at temperatures above 500 K. The elongation of $\tau_{avr}$ at high temperatures provides additional confirmation of the $\beta \rightarrow \alpha$ phase transition. Moreover, the initial $\tau_{avr}$ values at 83 K decrease with increasing $Eu^{3+}$ ions concentration, as expected from the previously described mechanism. In contrast, for 2% $Eu^{3+}$, an increase in temperature from 200 K to 550 K leads to a slight reduction in $\tau_{avr}$, with further shortening observed above 550 K. To quantitatively characterize these thermal variations, the $S_R$ was determined in a manner analogous to the *LIR*. For reliable thermometric measurements, it is crucial that the parameter, in this case $\tau_{avr}$, exhibits monotonic behavior over the analyzed temperature range. The observed nonmonotonic behavior, where $\tau_{avr}$ decreases then increases with temperature, necessitated evaluating $S_R$ in two separate thermal ranges: Range 1 (83 – 200 K) and Range 2 (200 – 550 K). The rapid decrease in $\tau_{avr}$ at the $\gamma_{HT} \rightarrow \alpha$ phase transition yields a maximum $S_{R1max}$ of 0.38% $K^{-1}$ for 0.2% $Eu^{3+}$ (Figure 7b). Increasing the $Eu^{3+}$ ions concentration results in a reduction of the relative sensitivity and a monotonic shift of the temperature at which $S_{R1max}$ is observed (Figure 7c), directly correlating with the change in the phase transition temperature. Although Range 2 is wider than Range 1, the $S_R$ values in this range are significantly lower (approximately 0.02% $K^{-1}$ at about 420 K) due to the gradual elongation of $\tau_{avr}$ (Figure 7d). These sensitivity values render the sensor impractical for thermometric applications in Range 2, leading to the conclusion that $Li_3Sc_2(PO_4)_3:Eu^{3+}$ can be effectively utilized as a lifetime-based thermometer only in the temperature range below 200 K.

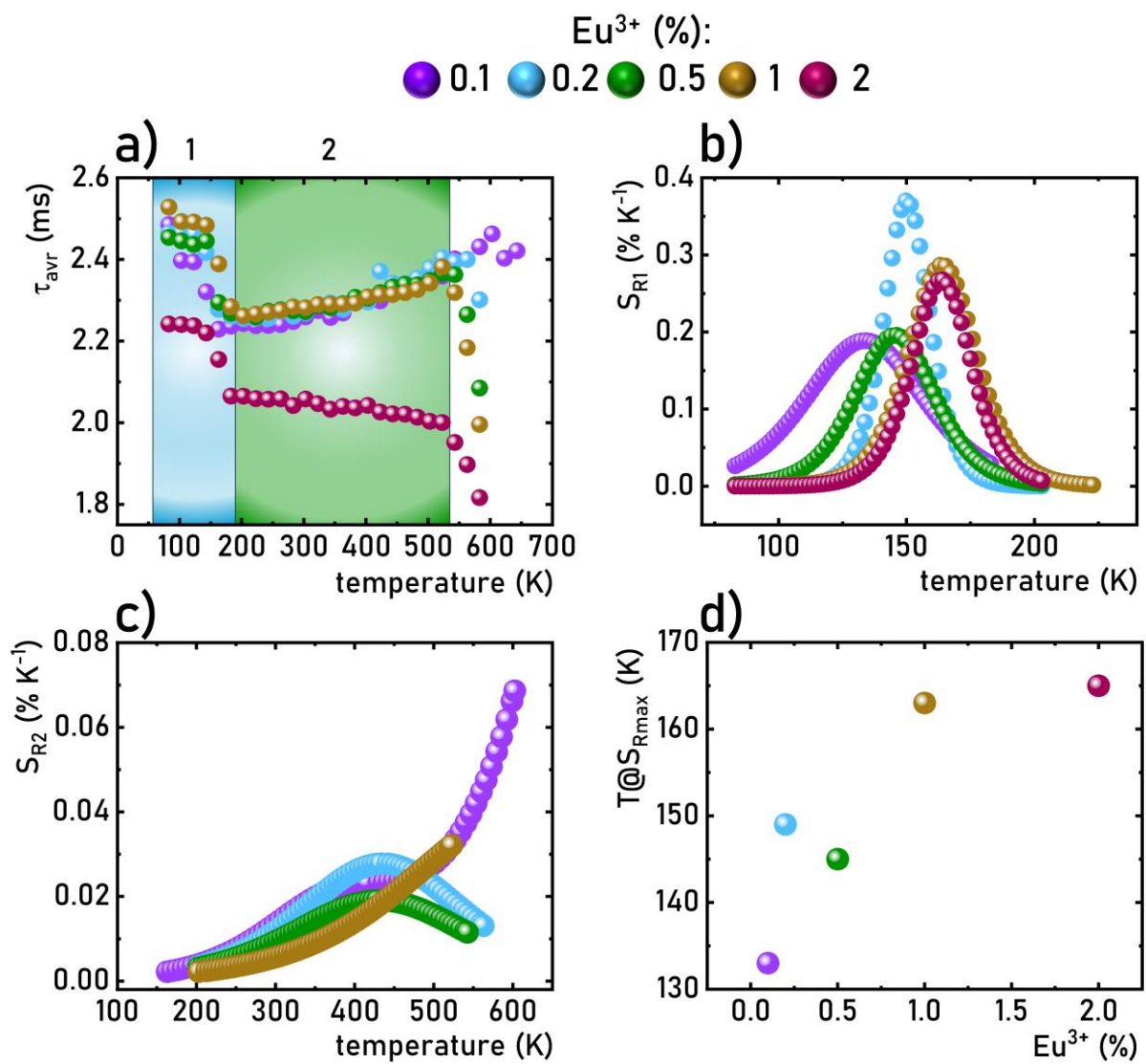

**Figure 7.** Thermal dependence of $\tau_{avr}$ of the $^5D_0$ state of $Eu^{3+}$ ions for different dopant concentration -a) and corresponding $S_R$ determined in the Range 1 -b) the influence of $Eu^{3+}$ ions concentration on the $T@S_{R1max}$-c) the thermal dependence of $S_R$ determined in Range 2 -d).

As is well known, the size of a crystallite affects the phase transition temperature which is particularly evident for small crystallites[41–45]. This effect is primarily related to the stresses arising in the crystal bounded by its surface. In such a case, the phase transition temperature between low symmetry and high symmetry is expected to increase with a particle size while for the reverse transition a reduction in the phase transition temperature is expected. On the other hand, it is well known that an increase in the annealing temperature creates favorable thermodynamic conditions for crystal growth leading to their expansion[46–48]. Considering that the phase-transition temperature affects the operating range of such a luminescent thermometer, the effect of annealing temperature on the thermometric performance of a luminescent thermometer based on $Li_3Sc_2(PO_4)_3$:$Eu^{3+}$ was verified. Therefore, the $Li_3Sc_2(PO_4)_3$:$Eu^{3+}$ badge was divided into 5 parts and annealed at different temperatures in the range of 1173-1573 K while keeping the annealing time constant. The recorded representative SEM images for such samples show significant morphological differences of such materials (Figure 8a-e). Calculations of average crystallite size show a clear and monotonic increase in crystallite size from 4.8 μm for 1173 K annealing temperature to 22.7 μm for 1573K (Figure 8f, S48-S52). On the other hand, DSC studies indicate that the temperature of the $\gamma_{LT} \rightarrow \alpha/\beta$ phase transition successively increases from 165 K for particle size 4.8 μm to 175 K for particles of 17.2 μm and then decreases to 159 K for particles of 22.7 μm annealed at 1573 K (Figure 8g). An analogous but reversed trend is observed for the $\alpha/\beta \rightarrow \gamma_{HT}$ transition where a decrease from 540 K for 4.8 μm crystals to 490 K for 17.2 μm crystals is observed followed by an increase to 510 K for 22.7 μm. This effect is most likely related to a change in the volume of the elemental unit cell which increases monotonically with increasing annealing temperature up to 1473 K, while further increases in temperature result in a cell volume decrease. Although, the reason for this decrease in unit cell volume for high annealing temperatures is not clear it can be confirmed that the phase transition temperature can be controlled by changing the morphology of $Li_3Sc_2(PO_4)_3$:$Eu^{3+}$. The thermal dependence of $LIR_1$ for these phosphors indicates important information (Figure 8h). First, the value of $LIR_1$ at 83 K shows a similar relationship to the phase transition temperature, that is, it increases with increasing annealing temperature and then decreases for crystals annealed at 1573 K to a value comparable to that obtained for an annealing temperature of 1073 K. The increase in the measurement temperature results, for all the materials analyzed, in a similar relationship to the one described earlier, but the dynamics of the observed changes decreases as the annealing temperature increases. Similarly, in the case of $LIR_2$, the effect of annealing temperature on the dynamics of changes in the thermometric parameter is observed, which is reflected in the $S_R$ values obtained for these materials (Figure 8i). The maximum value of $S_R$ for crystallites with a size of 4.8 µm was 2.0% $K^{-1}$ and increased successively with increasing crystallite size reaching a maximum for $Li_3Sc_2(PO_4)_3:Eu^{3+}$ annealed at 1273 K amounting to $S_{Rmax}$=7.8% $K^{-1}$ and then the value of $S_R$ successively decreased with further increase in annealing temperature obtaining 1.2% $K^{-1}$ for 1573 K. In addition, it is worth noting that the temperatures at which $S_{Rmax}$ was obtained reflect the changes in phase transition temperature obtained from DSC analysis.

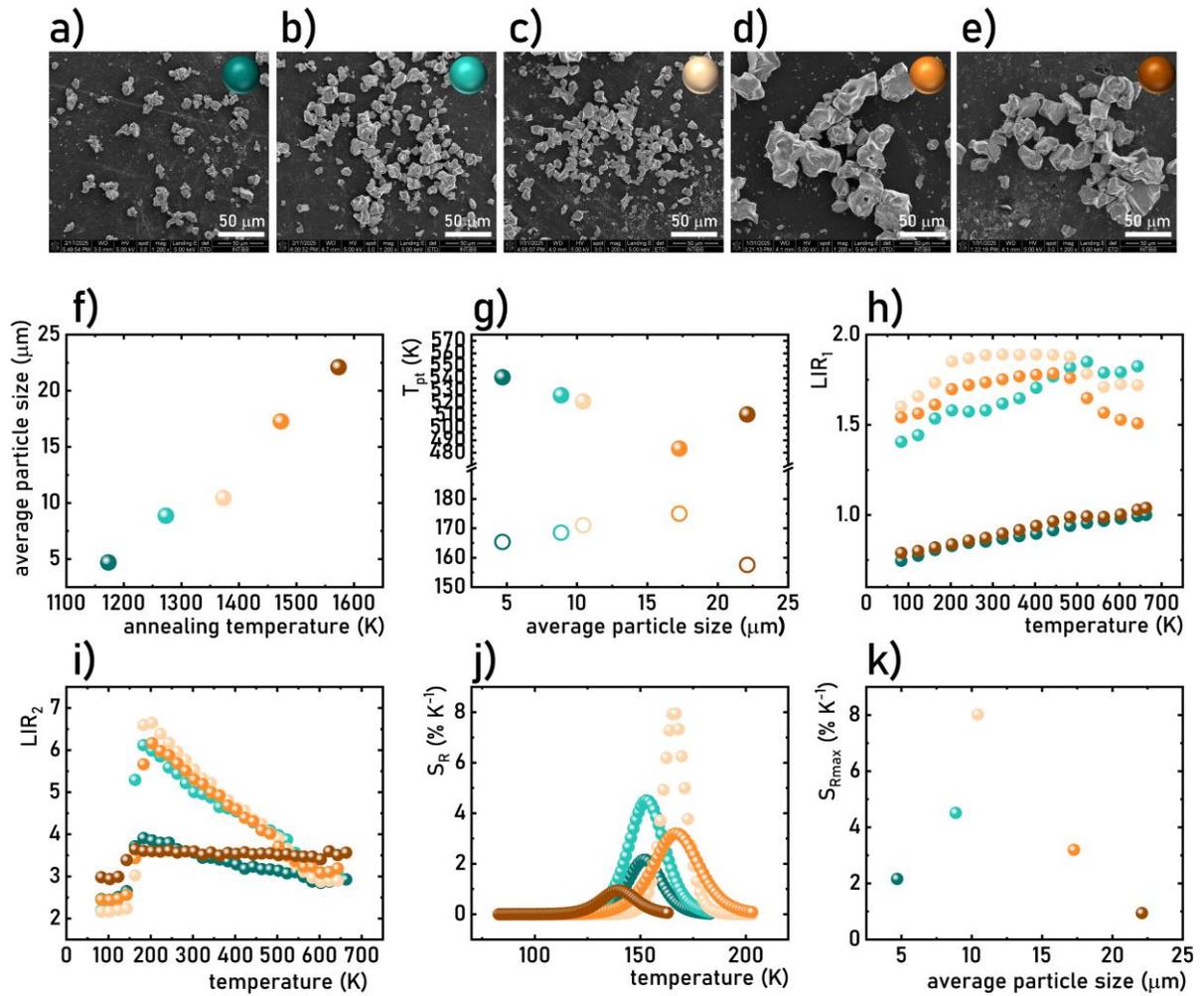

**Figure 8.** Representative SEM images for $Li_3Sc_2(PO_4)_3$:1%$Eu^{3+}$ annealed at 1173 K -a); 1273 K-b); 1373 K -c) 1473 K -d) and 1573 K -e), the influence of the annealing temperature on the average particle size -f); the influence of the particle size of $Li_3Sc_2(PO_4)_3$:1%$Eu^{3+}$ on the $\gamma_{LT}\rightarrow\alpha/\beta$ (empty dots) and $\alpha/\beta\rightarrow\gamma$ (full dots) phase transition temperature-g) thermal dependence of $LIR_1$-h) and $LIR_2$-i) for $Li_3Sc_2(PO_4)_3$:1%$Eu^{3+}$ annealed at different temperatures; and corresponding $S_R$ for $LIR_2$ -j); the influence of the average particle size on the $S_{Rmax}$- k).

## Conclusions

In this study, we investigated the spectroscopic properties of $Li_3Sc_2(PO_4)_3$:$Eu^{3+}$ over a broad temperature range to assess its potential for luminescence thermometry based on structural phase transitions. Differential scanning calorimetry and *in-situ* temperature-dependent X-ray diffraction analyses revealed two distinct structural phase transitions: a known monoclinic α/β to orthorhombic $\gamma_{HT}$ transition at approximately 550 K, and a previously unreported orthorhombic $\gamma_{LT}$ to monoclinic α/β transition at around 160 K. These transitions significantly alter the local symmetry of the crystallographic sites occupied by $Eu^{3+}$ ions, leading to notable modifications in their spectroscopic behavior, thereby enabling their application in luminescent thermometry. The $Eu^{3+}$ ions in monoclinic $Li_3Sc_2(PO_4)_3$ occupy two distinct $Sc^{3+}$ sites, denoted as *A* and *B*, each exhibiting unique luminescence characteristics. At low dopant concentrations, $Eu^{3+}$ ions preferentially occupy site *A*. As the $Eu^{3+}$ ions concentration increases to 1%, occupancy of site *B* begins to rise gradually. Notably, at a concentration of 2% $Eu^{3+}$, there is an abrupt, nearly twelvefold enhancement in the emission intensity from site *B*. Analysis of the temperature dependence of the $LIR_2$ identified two temperature intervals exhibiting monotonic, albeit opposite, $LIR_2$ variations, corresponding to the structural phase transitions. Each interval represents a distinct operational range for the luminescent thermometer. The highest relative sensitivity was observed at 148 K for 0.1% $Eu^{3+}$, with $S_R = 7.8\%$ $K^{-1}$. Increasing the $Eu^{3+}$ concentration resulted in a decrease in $S_R$ and an upward shift in the temperature at which $S_{Rmax}$ occurs. In the higher temperature range, a maximum $S_R$ of 0.65% $K^{-1}$ was recorded at 550 K for 0.5% $Eu^{3+}$, with higher dopant concentrations leading to a gradual decrease in the phase transition temperature. Luminescence lifetime measurements indicated an abrupt shortening in the $\tau_{avr}$ at the $\gamma_{LT} \to \alpha$ phase transition temperature, followed by a gradual increase with rising temperature. However, the relative sensitivities associated with these lifetime changes were relatively low, peaking at 0.39% $K^{-1}$ at 151 K for 0.2% $Eu^{3+}$. Furthermore, increasing the annealing temperature of $Li_3Sc_2(PO_4)_3$:$Eu^{3+}$ led to an increase in average crystallite size, accompanied by a gradual expansion of the unit cell volume up to 1373 K, beyond which a decrease in volume was observed with further temperature elevation. This behavior correlates directly with variations in relative sensitivity and the temperature corresponding to $S_{Rmax}$. To our knowledge, Li$_3$Sc$_2$(PO$_4$)$_3$:Eu$^{3+}$ represents the first luminescent thermometer based on first-order structural phase transitions that operates effectively across two distinct temperature ranges. Moreover, we demonstrate that the thermometric performance of this material can be fine-tuned by adjusting both the dopant ion concentration and the annealing temperature of the phosphor.

**Acknowledgements**

This work was supported by the National Science Center (NCN) Poland under project no. DEC-UMO- 2022/45/B/ST5/01629.